\documentstyle[aps,preprint]{revtex} 
\begin{document} 
\draft 
\preprint{IMSc-97/11/39} 
\title{Quantum Black Hole Entropy}
\author{Romesh K. Kaul and Parthasarathi Majumdar\footnote{email: kaul, partha,  
@imsc.ernet.in}} 
\address{The Institute of Mathematical Sciences, CIT Campus, Madras 600113, India. }
\maketitle 
\begin{abstract}
We derive an exact formula for the dimensionality of the Hilbert space of the boundary 
states of $SU(2)$ Chern-Simons theory, which, according to the 
recent work of Ashtekar {\it et al}, leads to   
the Bekenstein-Hawking entropy of a four dimensional Schwarzschild black hole. Our
result stems from the relation between the (boundary) Hilbert space of the Chern-Simons 
theory  with the space of conformal blocks of the Wess-Zumino model on the boundary
2-sphere.
\end{abstract}
\vglue .2in

The issue of the Bekenstein-Hawking (B-H) \cite{bek}, \cite{haw}  entropy of black holes
has been under
intensive scrutiny for the last couple of years, following the derivation of the entropy
of certain extremal charged black
hole solutions of toroidally compactified heterotic string and also type
IIB superstring from the underlying string theories \cite{sen},
\cite{strv}. In the former case of the heterotic string, the entropy was shown to be
proportional to the area of the `stretched' horizon of the corresponding extremal
black hole, while in the latter case it turned out to be {\it precisely} the
B-H result. The latter result was soon generalized to a large number
of four and five dimensional black holes of type II string theory and M-theory (see
\cite{hor} for a review), all of which could be realized as certain D-brane
configurations and hence saturated the BPS bound. Unfortunately, the simplest black
hole of all, the four dimensional Schwarzschild black hole, does not appear to be
describable in terms of BPS saturating D-brane configurations, and hence is not
seemingly amenable to such a simple analysis.\footnote{Recently, Sfetsos and
Skenderis, hep-th/9711138, have obtained the B-H result for the 4d Schwarzschild black
hole by a U-duality map to the 3d BTZ black hole whose entropy has been
calculated, again for large areas, by Carlip\cite{bal}. See also, Arguiro
{\it et al}, hep-th/9801053.} Results for near-extremal black holes, modeled as
near-BPS states of string
and M-theory, have also been obtained \cite{hor}, pertaining both to their entropy and
also Hawking radiation. In the majority of the cases considered, complicated (sometimes
intersecting) configurations of D-branes were treated in the `effective string
approximation' \cite{dm}; in this approximation, computations effectively reduce to
that in a two dimensional conformal field theory \cite{str}. 

The B-H entropy of a four dimensional Schwarzschild black hole 
has been obtained, for large areas of the event horizon, within the
alternative framework of canonical quantum gravity \cite{ash1} by Krasnov and Ashtekar
{\it et al} \cite{kras}, \cite{ash2},  up to an overall constant of $O(1)$ known as
the Immirzi 
parameter \cite{imm} (which essentially characterizes inherent ambiguities in the
quantization scheme, and is therefore present in the quantum theory even in the
absence of black holes). The black hole spacetime is considered as a 4-fold 
bounded by 
the surface at asymptotic null infinity (on which standard asymptotically flat boundary 
conditions hold) and the event horizon (on which boundary conditions, special to the 
spherically symmetric Schwarzschild geometry are assumed). The action embodying the 
assumed boundary conditions consists of, over and above the Einstein-Hilbert action (in 
the Ashtekar variables \cite{ash1}), an $SU(2)$ Chern-Simons (CS) gauge theory 
`living' on a 
coordinate chart of constant finite cross sectional area $A_S$ (and possessing some other 
properties) on the 
horizon. The Chern-Simons coupling parameter $k \sim A_S$. In the Hamiltonian formulation of 
the theory, the boundary conditions are implemented as a condition on the phase space 
variables 
restricted to the boundary (2-sphere) of a spacelike 3-surface intersecting the constant 
area coordinate patch on the horizon. This results in a 
reducible connection variable which is gauge fixed (on the boundary) to the $U(1)$ 
subgroup of the $SU(2)$ invariance of the CS theory. In the quantum theory, the boundary 
conditions, implemented as an operator equation, imply that the space of surface 
(boundary) quantum  states is composed of subspaces given by the Hilbert 
space of the $U(1)$ CS theory, on the boundary 2-sphere {\it with finitely many 
punctures $p$ labeled by spins $j_p$}. Now, it has been argued \cite{bal} that boundary 
(or `edge') states play the major 
role in producing black hole entropy. Likewise, the entropy of the black hole 
under consideration is assumed to 
emerge only from the surface states, and defined by tracing over the `volume' states. It
is then given by  $S_{bh} = ln N_{bh}$ where $N_{bh}$ is the number of boundary 
CS states. This number is next obtained from the dimensionality of the Hilbert space of 
boundary $U(1)$ CS states on the punctured 2-sphere. Finally, this is compared with the 
spectrum \cite{rov} of the area operator \cite{ashle} in canonical quantum gravity,
known, upto the Immirzi parameter, in terms of spins $j_p$ on the punctures $p$. For 
large number of punctures and large area, it is seen to be proportional to 
the logarithm of the dimensionality of the space of boundary CS states (i.e., the entropy). A 
particular choice of the Immirzi parameter then reproduces the Bekenstein Hawking value.

Recall that the B-H entropy of a black hole was proposed on the basis of 
semiclassical analyses, and as such, is by no means beyond 
modification in the full quantum theory. One of the simplifying steps in the above 
derivation was 
the reduction of the gauge group of the CS theory from  the original $SU(2)$ to $U(1)$, 
by gauge fixing the connection on the boundary 2-sphere. As admitted by the authors, this 
is not a necessary step. Indeed, there exist powerful results relating the state space 
of CS theories on 3-folds with boundary
to the conformal blocks of an $SU(2)$ Wess-Zumino model of level $k$ on that boundary \cite{wit}. It stands to reason that the entropy, derived from such considerations, 
will be more exact quantum mechanically. In this paper, we focus on such a calculation
of the entropy, using the formalism of two dimensional conformal field theory.

More specifically, we compute the number of conformal blocks of an 
$ SU(2)_k $ Wess-Zumino theory
on a  punctured 2-sphere, for a set of punctures ${\cal P}\equiv \{1,2, \dots, p\}$
where these 
punctures are labeled by the spin $j_p$, {\it for arbitrary level $k$ (corresponding
to an arbitrary area of the 
cross section of the patch chosen on the horizon)}.\footnote{Similar ideas using
self-dual boundary conditions as outer boundary conditions have appeared in
\cite{smol1}.} This number
can be computed in terms of the so-called fusion  matrices $N_{ij}^{~~r}$ \cite{dms} 
\begin{equation}
N^{\cal P}~=~~\sum_{\{r_i\}}~N_{j_1 j_2}^{~~~~r_1}~ N_{r_1 j_3}^{~~~~r_2}~ N_{r_2
j_4}^{~~~~r_3}~\dots  \dots~ N_{r_{p-2} 
j_{p-1}}^{~~~~~~~~j_p} ~ \label{fun} \end{equation}
Here, each matrix element $N_{ij}^{~~r}$ is $1 ~or~ 0$, depending on whether the
primary field  $[\phi_r]$ is allowed  or not in the conformal field theory
fusion algebra for the primary fields $[\phi_i]$ and $[\phi_j] $ ~~($i,j,r~ =~ 0,
1/2, 1, ....k/2$):
\begin{equation}
[\phi_i] ~ \otimes~ [\phi_j]~=~~\sum_r~N_{ij}^{~~r} [\phi_r]~ . \label{fusal}
\end{equation}
Eq. (\ref {fun} ) gives the number of conformal blocks with spins $j_1, j_2, \dots,
j_p$ on $p$ external lines and spins $r_1, r_2, \dots, r_{p-2}$ on
the internal lines. We next take 
recourse to the Verlinde formula \cite{dms}
\begin{equation}
N_{ij}^{~~r}~=~\sum_s~{{S_{is} S_{js} S_s^{\dagger r }} \over S_{0s}}~, \label{verl}
\end{equation}
where, the unitary matrix $S_{ij}$ diagonalizes the fusion  matrix. Upon using the 
unitarity of the $S$-matrix, the algebra (\ref{fun}) reduces to 
\begin{equation}
N^{\cal P}~=~ \sum_{r=0}^{k/2}~{{S_{j_1~r} S_{j_2~r} \dots S_{j_p~r}} \over (S_{0r})^{p-2}}~.
\label{red} \end{equation}
Now, the matrix elements of $S_{ij}$ are known for the case under 
consideration ($SU(2)_k$ Wess-Zumino model); they are given by
\begin{equation}
S_{ij}~=~\sqrt{\frac2{k+2}}~sin \left({{(2i+1)(2j+1) \pi} \over k+2} \right )~, \label{smatr}
\end{equation}
where, $i,~j$ are the spin labels, $i,~j ~=~ 0, 1/2, 1,  .... k/2$. Using this $S$-matrix, the number of conformal blocks for the set of 
punctures ${\cal P}$ is given by
\begin{equation}
N^{\cal P}~=~{2 \over {k+2}}~\sum_{r=0}^{ k/2}~{ {\prod_{l=1}^p sin \left( 
{{(2j_l+1)(2r+1) \pi}\over k+2} \right) } \over {\left[ sin \left( {(2r+1) \pi \over k+2} 
\right)\right]^{p-2} }} ~. \label{enpi} \end{equation}
In the notation of \cite{ash2}, eq. (\ref{enpi}) gives the dimensionality, $dim ~{\cal 
H}^{\cal P}_S$, {\it for arbitrary area of the horizon $k$ and arbitrary number of 
punctures}. The dimensionality of the space of states ${\cal H_S}$ of CS theory on three-manifold with $S^2$ boundary   is then given 
by summing $N^{\cal P}$ over all sets of punctures ${\cal P}:
~ 
N_{bh}~=~\sum_{\cal P} N^{\cal P}$. Then, the entropy of the black hole is given by 
$S~=~\log N_{bh}$.

Observe now that eq. (\ref{enpi}) can be rewritten, with appropriate redefinition of 
dummy variables and recognizing that the product can be written as a multiple sum,
\begin{equation}
N^{\cal P}~=~\left ( 2 \over {k+2} \right) ~\sum_{l=1}^{k+1} sin^2 
\theta_l~\sum_{m_1 = 
-j_1}^{j_1} \cdots \sum_{m_p=-j_p}^{j_p} \exp \{ 2i(\sum_{n=1}^p m_n)~ \theta_l \}~, 
\label{summ} \end{equation}
where, $\theta_l ~\equiv~ \pi l /(k+2)$. Expanding the $\sin^2 \theta_l$ and 
interchanging the order of the summations, a few manipulations then yield
\begin{equation}
N^{\cal P}~=~\sum_{m_1= -j_1}^{j_1} \cdots \sum_{m_p=-j_p}^{j_p} \left[ 
~\delta_{(\sum_{n=1}^p m_n), 0}~-~\frac12~ \delta_{(\sum_{n=1}^p m_n), 1}~-~ 
\frac12 ~\delta_{(\sum_{n=1}^p m_n), -1} ~\right ], \label{exct}
\end{equation}
where, we have used the standard resolution of the periodic Kronecker deltas in terms of 
exponentials with period $k+2$,
\begin{equation}
\delta_{(\sum_{n=1}^p m_n), m}~=~ \left( 1 \over {k+2} \right)~ \sum_{l=0}^{k+1} \exp 
\{2i~[ (\sum_{n=1}^p m_n)~-~m] \theta_l \}~. \label{resol}
\end{equation}
Notice that the explicit dependence on $k+2$ is no longer present in the exact formula 
(\ref{exct}). 

For large $k$ and large number of punctures $p$ our result (\ref{enpi}) reduces to 
\begin{equation}
N^{\cal P}~~\sim~~\prod_{l=1}^p~(2j_l~+~1)~~\label{bigk}
\end{equation}
in agreement with the result of ref. \cite{ash2}. Thus the semiclassical B-H formula
is valid in this approximation. To see if the B-H formula 
relating entropy with area is valid even in the
quantum theory, one needs to obtain the eigenvalues of the area operator without any
assumptions about their size. This might entail a modified regularized operator which
measures horizon area in the quantum theory and is, in general, a constant of motion,
i.e., commutes with the Hamiltonian constraint. 

It appears that methods of two dimensional conformal field theory
effectively describe {\it quantitative} quantum physics of
the black holes in four
spacetime dimensions. In our work, the conformal field theory enters
through the relation the boundary states of the $SU(2)$ Chern-Simons theory have with the conformal blocks of the corresponding
conformal field theory. The theory of irreducible representations 
of the simplest of the conformal field theories,
$SU(2)_k$ Wess-Zumino model, is crucial to yield what
may be thought of as a quantum generalizations of the semiclassical B-H entropy of the black hole.
Extensions of our results to the case of charged and rotating black holes
will hopefully constitute a future publication, as also attempts to understand Hawking
radiation within the canonical quantum gravity approach.

Discussions with S. Carlip, T. R. Govindarajan and C. Rovelli are gratefully acknowledged.

\end{document}